\documentclass[
reprint,
superscriptaddress,
amsmath,amssymb,
aps,
prl,
]{revtex4-2}

\usepackage{float}
\usepackage{graphicx}
\usepackage{dcolumn}
\usepackage{bm}
\usepackage[colorlinks,linkcolor=blue,anchorcolor=blue,citecolor=blue,urlcolor=blue]{hyperref}
\usepackage{longtable}

\begin{document}

\title{Phase-locked phonon laser enhanced ultra-weak force measurement}

\author{Yu Zheng}
\thanks{These authors contributed equally to this work.}
\affiliation{Laboratory of Quantum Information, CAS Center for Excellence in Quantum Information and Quantum Physics, University of Science and Technology of China, Hefei 230026, China}
\author{Long Wang}
\thanks{These authors contributed equally to this work.}
\affiliation{Laboratory of Quantum Information, CAS Center for Excellence in Quantum Information and Quantum Physics, University of Science and Technology of China, Hefei 230026, China}
\author{Lyu-Hang Liu}
\affiliation{Laboratory of Quantum Information, CAS Center for Excellence in Quantum Information and Quantum Physics, University of Science and Technology of China, Hefei 230026, China}
\author{Yuan Tian}
\affiliation{Laboratory of Quantum Information, CAS Center for Excellence in Quantum Information and Quantum Physics, University of Science and Technology of China, Hefei 230026, China}
\author{Xiang-Dong Chen}
\affiliation{Laboratory of Quantum Information, CAS Center for Excellence in Quantum Information and Quantum Physics, University of Science and Technology of China, Hefei 230026, China}
\affiliation{Hefei National Laboratory, University of Science and Technology of China,
Hefei 230088, China}
\author{Dong Wu}
\affiliation{State Key Laboratory of Opto-Electronic Information Acquisition and Protection, Key Laboratory of Precision Scientific Instrumentation of Anhui Higher Education Institutes, Department of Precision Machinery and Precision instrumentation, University of Science and technology of China, Hefei 230026, China}

\author{Guang-Can Guo}
\author{Fang-Wen Sun}
 \email{fwsun@ustc.edu.cn}
\affiliation{Laboratory of Quantum Information, CAS Center for Excellence in Quantum Information and Quantum Physics, University of Science and Technology of China, Hefei 230026, China}
\affiliation{Hefei National Laboratory, University of Science and Technology of China,
Hefei 230088, China}

\date{\today}

\begin{abstract}
Optically levitated micro- and nanoparticles are an ideal optomechanical platform for precision measurements, particularly enabling the detection of ultraweak forces. Nevertheless, quantum backaction and inherent instabilities induced by the trapping laser fundamentally restrict further improvements in force sensitivity and resolution. To circumvent these bottlenecks, we actively drive the levitated nanoparticle's mechanical motion in a phase-locked phonon laser mode and integrate a carrier-modulation measurement architecture to enhance force sensing capabilities.  
The stable and high-amplitude oscillation of the phonon laser allows for the robust trapping under $1$ $\mathrm{ mW}$-level laser power, which in turn reduces the force noise to $4.0(3)\times10^{-22}\mathrm{~N/Hz^{1/2}}$.  
Furthermore, by using phase-locked phonon laser, the measurement system achieves active stabilization and extended coherence time with the measured signal to $12,500$ seconds, realizing a measurement resolution of $8(4)\times10^{-24}$ N with a sensitivity of $9.3(7)\times10^{-22} \mathrm{N/Hz^{1/2}}$ under a loaded force. These results establish the phonon laser as a low-noise, long-coherence-time, self-stabilizing platform for precision measurements, as well as in quantum and fundamental physics tests. 
\end{abstract}

\maketitle
Optically levitated micro- and nanoparticles in vacuum have emerged as versatile platforms for frontier physics \cite{review2021sci}. They have enabled groundbreaking research into macroscopic quantum mechanics \cite{Gcool2020sci,tebbenjohanns2021quantum,magrini2021real,dania2025high} and precision measurements of physical quantities including torques \cite{ahn2020ultrasensitive}, masses \cite{ricci2019accurate,zheng2020robust}, charges \cite{moore2014search,frimmer2017controlling}, accelerations \cite{monteiro2017optical}, and gas pressures \cite{blakemore2020absolute,liu2024nanoscale}. Fundamentally, these sensing capabilities rely on detecting the forces arising from the respective interactions \cite{ranjit2015attonewton,hempston2017force,hebestreit2018sensing,liang2023yoctonewton}. Therefore, pushing the boundaries of force sensing directly enhances the discovery potential of levitated systems in areas such as the search for dark matter \cite{monteiro2020darkmatter,kilian2024dark}, high-frequency gravitational waves \cite{arvanitaki2013gwave,winstone2022gwave}, short-range interactions \cite{geraci2010short,kawasaki2020high}, and non-Newtonian gravitational effects \cite{geraci2010short,blakemore2021nonnewton}. 

However, further enhancement of the force measurement performance in vacuum optical levitation systems is still constrained by two limiting factors, which are force noise and system instability.
The force noise, originating from gas-molecule collisions \cite{li2012fundamental} and trapping laser \cite{jain2016recoil,Savard1997noise}, limits the measurement precision per unit time, i.e., measurement sensitivity.
The system instability, caused by factors like laser intensity fluctuations, leads to time-varying parameters of the levitated oscillator, particularly the eigen-frequency. This compromises long-term measurement reliability and thus limits the achievable force resolution.

Moreover, in many mechanical sensors, including self-oscillating systems, the detected signals such as weak forces are inferred from changes in oscillator parameters such as frequency and amplitude \cite{pl2019pettit,pl2023zheng,pl2023Kuang,pl2024xiao,pan2024ultra,GUha2020force,Javid2021force,Liu2021Phonon}. However, these same observables are also susceptible to noise from drift and intrinsic fluctuations, so signal and noise are encoded in the same channel. As a result, experimental control based on the same parameters used for signal readout can hardly be applied to enhance the sensitivity and resolution.

\begin{figure*}[t]
    \centering
    \includegraphics[width=\linewidth]{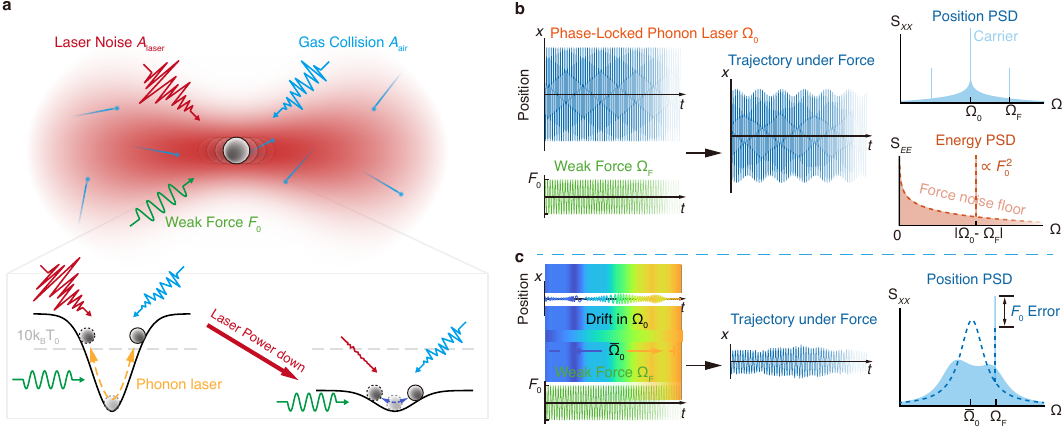}
    \caption{\label{fig:1}Illustration of two-step weak force sensing enhanced by a phonon laser. (a) Step 1: Low-power trapping for force-noise suppression. An optically levitated nanoparticle subjected to a weak force $F_0$ also experiences stochastic forces from residual-gas collisions, $A_\text{air}$, and laser induced force noise, $A_\text{laser}$. Lowering the trapping power suppresses laser-related noise but also reduces the trap depth. The phonon-laser state enables stable trapping during power reduction.
(b) Step 2: Carrier-modulation force sensing with a phase-locked phonon laser (PLPL). A sinusoidal weak force generates sidebands around the stable phonon-laser carrier. The resulting periodic modulation of the oscillator energy gives rise to a sharp peak in the energy power spectral density (PSD), with a height proportional to the square of the force. The phase-locking stabilizes the system, resulting in agreement between the measured data and the theoretical prediction (dashed line). (c) Comparison: Force sensing with centre-of-mass (COM) cooling in an unstable system. The motion of the COM cooled oscillator is driven by the loaded weak force. Due to fluctuations in the particle's eigenfrequency, the measured PSD signal no longer matches the ideal prediction (dashed line), leading to force measurement errors.}
\end{figure*}

Here we introduce a two-step strategy for ultra-weak force sensing based on a phonon laser, as shown in Fig. \ref{fig:1}. In the first step, the high and stable oscillation amplitude of the phonon laser provides a high signal-to-noise ratio (SNR) readout, allowing us to reduce the trapping laser power by two orders of magnitude to 1 mW, rendering laser-induced noise negligible and thereby achieving a force noise of $4.0(3)\times10^{-22}\,\mathrm{N/Hz^{1/2}}$. In the second step, we introduce a ``carrier-modulation'' measurement architecture for force sensing, in which weak forces are encoded in sidebands around a stable phase-locked phonon-laser (PLPL) carrier rather than in shifts of the mechanical eigenmode. This separation decouples signal transduction from oscillator stabilization. In our implementation, the carrier is phase-locked to an external reference clock through a phase-locked loop (PLL), suppressing environmental drift while preserving the force-induced response. By combining low-power operation with PLPL, the system simultaneously suppresses laser-induced force noise and compensates for long-term instability. This extends the optimal averaging time to 12,500 s, yielding a force resolution of $8(4)\times10^{-24}\,\mathrm{N}$ with a sub-zeptonewton sensitivity of $9.3(7)\times10^{-22}\,\mathrm{N}/\sqrt{\mathrm{Hz}}$. These results push the force-sensing performance of optomechanical systems into a new regime and establish a general strategy for improving measurement performance. In addition, operation at ultra-low optical power provides a highly coherent, low-noise platform for macroscopic quantum state preparation and for levitating materials that are sensitive to laser heating.

\section{Results}

\subsection{Low-power trapping for noise suppression}

As shown in Fig. \ref{fig:1}(a), for an optically levitated oscillator subjected to a stochastic force noise, $F_{\text{sto}}(t) = A_{\text{sto}} \zeta(t)$ \cite{li2012fundamental}, where $\zeta(t)$ represents unit-variance white noise, the optimal force sensitivity is $S_\mathrm{F} = \sqrt{2} A_{\text{sto}}$.
The force noise can be decomposed as $A_{\text{sto}}^2 = A_{\text{air}}^2 + A_{\text{laser}}^2$. Here, $A_{\text{air}} = \sqrt{2k_\mathrm{B} T_0 \Gamma_0 m}$ comes from stochastic collisions with gas molecules, where $k_\mathrm{B}$ is the Boltzmann constant, $T_0$ is the environmental temperature, $m$ is the mass of the levitated particle, and $\Gamma_0$ is the air damping rate, which is proportional to the air pressure.

 The second noise contribution, $A_{\text{laser}}$, originates from the interaction between the levitated particle and the trapping laser. This noise term comprises several distinct mechanisms, including photon-recoil heating, feedback modulation induced noise, laser intensity and phase noise, and focal point instability. Despite the diverse physical origins of these contributions, they are all fundamentally related to the trapping laser power, $P$. The laser-induced noise $A_{\text{laser}}$ can be further divided into two main components. The first component scales as $A_{\text{laser}}^{(1)} \propto \sqrt{P}$ and is primarily associated with photon-recoil heating \cite{jain2016recoil}. The second component scales as $A_{\text{laser}}^{(2)} \propto P$ and mainly arises from focal-point shaking \cite{Savard1997noise}. Accordingly, the total laser noise can be written as
$A_{\text{laser}}^2 = (A_{\text{laser}}^{(1)})^2 + (A_{\text{laser}}^{(2)})^2$.
Therefore, the force sensitivity can be improved in four ways by reducing the particle mass, environmental temperature, air pressure, or laser power.

\begin{figure*}[t]
    \centering
    \includegraphics[width=\linewidth]{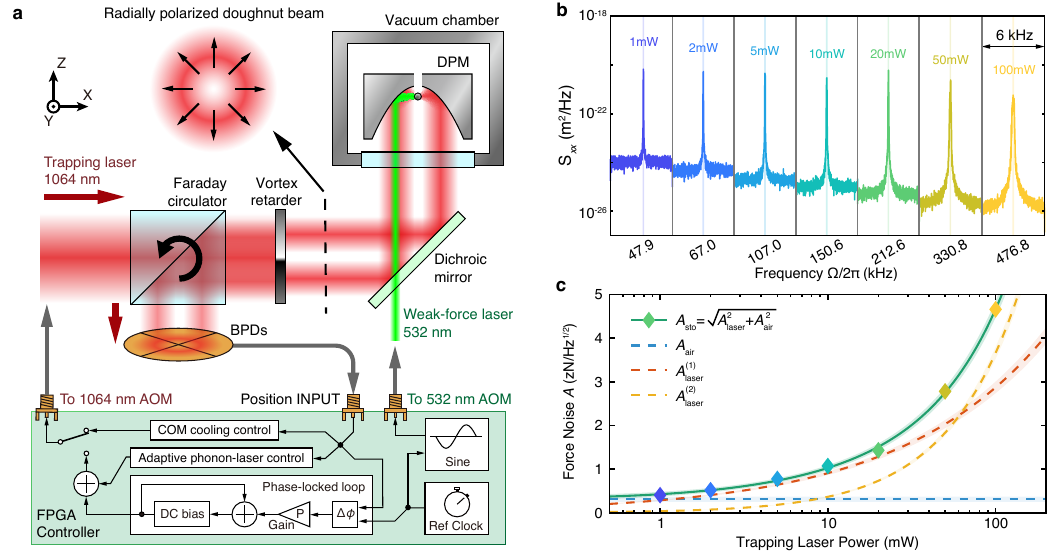}
    \caption{Experimental setup and trapping-power-dependent force noise. (a) A 1064-nm beam is converted to a radially polarized mode and focused by a deep parabolic mirror (DPM) to levitate a particle in vacuum. Motion is read out from DPM-collected scattered light, separated by a Faraday circulator and detected by balanced photodetectors (BPDs). A weak 532-nm laser, combined via a dichroic mirror, applies a weak optical force for measuring. The position signal is processed by an FPGA controller for phase-locked-loop operation, COM cooling and adaptive phonon–laser control via acousto-optic modulators (AOMs). (b) Position PSDs measured at different trapping powers with a cooling damp of $10\ \mathrm{Hz}$. (c) Measured force noise as a function of trapping power. Diamonds denote values extracted from the position PSDs. The curves show contributions from gas collisions, $A_{\mathrm{air}}$ (blue dashed), and laser-related noise, $A_{\mathrm{laser}}^{(1)}$ (red dashed) and $A_{\mathrm{laser}}^{(2)}$ (yellow dashed), together with the fitted total force noise, $A_{\mathrm{sto}}$ (green solid). The shaded region indicates the uncertainty of the fitted coefficients.
    \label{fig_low_power}}
\end{figure*}

However, as shown in Fig. \ref{fig:1}(a), reducing the power decreases the potential well depth. Once the well depth drops below $10k_\text{B}T_0$, a particle in thermal equilibrium can readily escape from it \cite{li2012fundamental}. To maintain stable levitation while reducing power, the mechanical energy of the levitated oscillator must be constrained. Centre-of-mass (COM) cooling, while commonly used for energy control \cite{cooling2011Li,paraCool2012}, is not an ideal method in this case because it is frequency-dependent. A decrease in laser power will lower the oscillator's frequency, causing a mismatch in frequency parameters and leading to cooling failure.

To overcome this challenge, we employ the phonon laser as the energy-constraining state. 
Under the phonon laser state, the oscillator's amplitude is stable, and its trajectory exhibits a high SNR periodic signal \cite{pl2019pettit,pl2023zheng}. By using simple rising-edge detection, the oscillator's phase and frequency can be accurately updated in real time, ensuring stable trapping during variations in laser power.

To verify this method, we experimentally constructed a vacuum optical levitation system, as shown in Fig. \ref{fig_low_power}(a). In this system, a 1064 nm laser beam is first converted into a doughnut-shaped, radially polarized mode and then focused by a deep parabolic mirror (DPM) to form the optical potential well \cite{Salakhutdinov2016DPM}. The light scattered by the nanoparticle is collected and retroreflected by the DPM, separated by a Faraday circulator, and detected by customized balanced photodetectors (BPDs) to monitor the particle’s spatial motion. The particle trajectory is recorded by a field-programmable gate array (FPGA)-based digital control system, which generates feedback signals for motion control.
Compared with conventional objective-based levitation systems, the DPM focuses the trapping beam from a nearly full $4\pi$ solid angle, yielding a tighter focus, stronger light--matter interaction and a deeper trapping potential, which enables trapping of smaller nanoparticles. We levitated a 90 nm-diameter silica nanoparticle for testing. Further details are provided in the Supplementary Information.

The generation of a phonon laser relies on a balance between linear gain (heating) and nonlinear dissipation (cooling) \cite{pl2019pettit}. It can be realized by deploying a feedback damping that depends on the phonon number, which is $\Gamma_{\mathrm{m}}(N) = \gamma_c N - \gamma_a$, where $N=E/\hbar\Omega_0$ is the phonon number, $E$ is the mechanical energy, $\hbar$ is the reduced Planck constant, $\Omega_0$ is the eigen-frequency of the oscillator, $\gamma_c$ is the nonlinear dissipation coefficient, and $\gamma_a$ is the linear gain coefficient. Once $\gamma_a$ exceeds the threshold, the levitated oscillator will enter the phonon laser state \cite{pl2023zheng}.

\begin{figure*}[t]
    \centering
    \includegraphics[width=\linewidth]{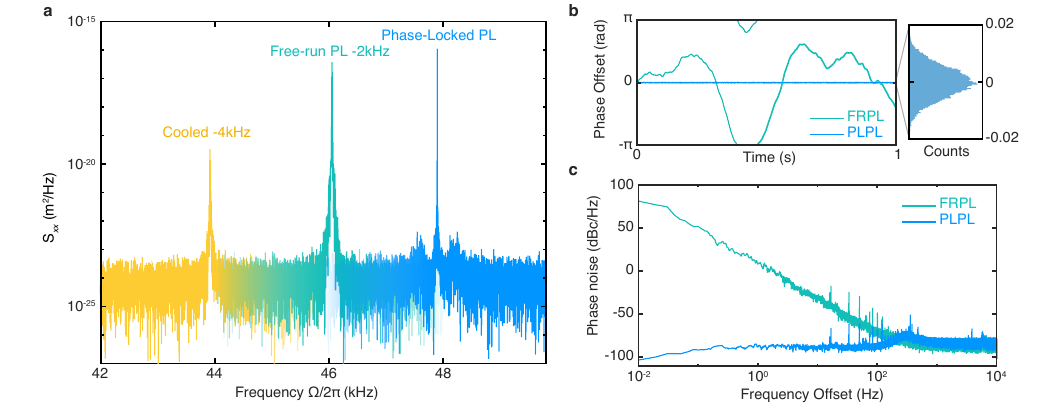}
    \caption{Characterization of the phase-locked phonon laser. (a) Position PSD for PLPL (blue line), FRPL (green line), and the COM cooling state (yellow line), respectively. The spectra are horizontally offset for clarity. (b) Phase offset time trace of PLPL and FRPL. The inset shows the histogram of the PLPL phase offset. (c) Phase noise PSD of the PLPL and the FRPL.
    \label{fig_PLPL}}
\end{figure*}

In the experiment, we progressively reduce the trapping power from $100$ mW to $1$ mW under the protection of the phonon laser (Supplementary Information). We pause at discrete levels and measure the system's force noise. Fig. \ref{fig_low_power}(b) shows that this power reduction decreases the X-axis oscillation frequency from $476.8$ kHz to $47.9$ kHz. For the force noise, the power reduction significantly suppresses laser relates noise $A_{\text{laser}}$. Owing to the high sensitivity of the DPM to beam pointing fluctuation induced aberrations, together with its tighter focus. the force noise at high trapping power is dominated by the focal instability related term $A_{\text{laser}}^{(2)} $. As the trapping power decreases, $A_{\text{laser}}^{(2)} $ rapidly diminishes, leaving the photon recoil related term $A_{\text{laser}}^{(1)} $ as the dominant laser noise contribution. When the trapping power reaches 1 mW, the force noise instead becomes dominated by the residual-air, $A_{\text{air}}$. At this point, the measured force noise is $4.0(3)\times10^{-22}\mathrm{~N/Hz^{1/2}}$, as shown in Fig. \ref{fig_low_power}(c). Uncertainties in all force-related quantities are obtained by propagating the uncertainty in the particle mass, with minor additional contributions from fitting errors (see Supplementary Information for details).

From the fitted $A_{\text{air}}$, we have an air equivalent pressure of $2.0(7)\times10^{-9}$ mbar for the system. The corresponding air damping rate is $\Gamma_0/2\pi=2.5(9)\times10^{-6} \mathrm{~Hz}$. which implies a quality factor of $Q=\Omega_\mathrm{0(1mW)}/\Gamma_0=1.9(7)\times10^{10}$ \cite{hq2024}.

\subsection{Force sensing with phase-locked phonon laser}

Since the particle's eigen-frequency is proportional to the square root of the trapping laser power ($\Omega_0\propto\sqrt{P}$), its frequency and phase can be rapidly modulated by actively controlling the laser power.
Drawing inspiration from the classic PLL design, a phase-locking module is added into the feedback control system.
The phase error between the phonon laser and a reference signal drives a feedback loop to maintain synchronization (as shown in Fig. \ref{fig_low_power}(a), see more details in Supplementary Information).

The results for phase locking are shown in Fig. \ref{fig_PLPL}. Compared to the free-run phonon laser (FRPL), the power spectral density (PSD) of PLPL exhibits a sharper peak at the locking frequency. The time trace of the phase offset of PLPL shows that its phase remains highly stable. The phase noise PSD in Fig. \ref{fig_PLPL}(c) shows that, compared to the FRPL, the PLPL achieves a dramatic reduction in low-frequency phase noise, with suppressions of 180 dB at 0.01 Hz and 100 dB at 1 Hz. The strong suppression of low-frequency noise indicates that the PLL is successfully stabilizing the system against frequency drift. In addition, the small noise bump observed around $300$ Hz is a “servo bump” caused by the feedback loop, which often appears in active laser frequency stabilization with Pound-Drever-Hall (PDH) technique \cite{bump2022Li}.

\begin{figure}[!htbp]
    \centering
    \includegraphics[width=\linewidth]{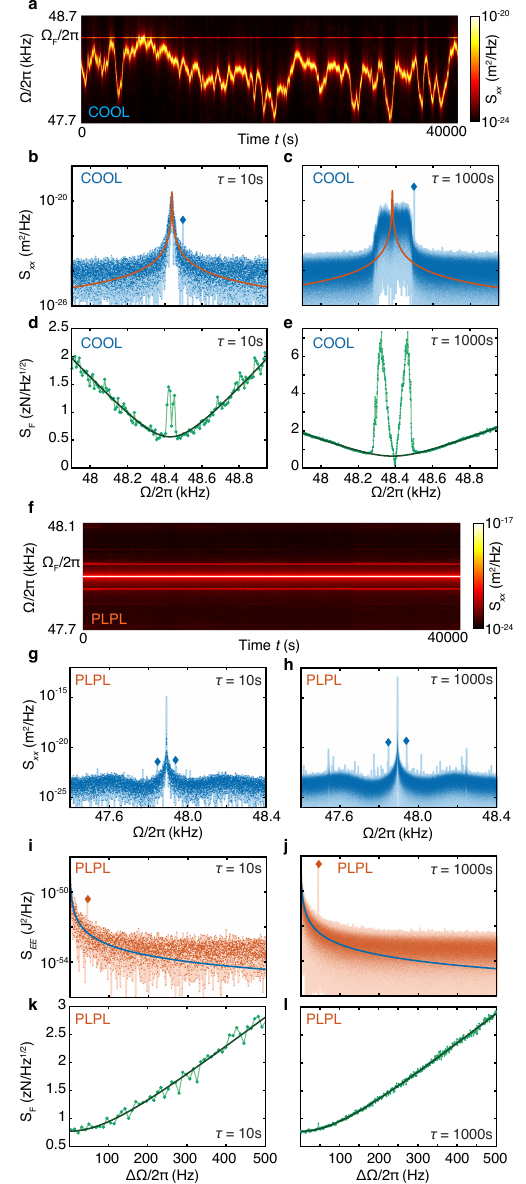}
    \caption{
Force measurement spectrum signals. (a) Time-resolved position PSD during force-sensing measurements under COM cooling. (b,c) Position PSDs in the  COM cooling state for sampling times of 10 and 1,000 s, respectively. (d,e) Corresponding force sensitivity spectra. (f) Time-resolved position PSD during force-sensing measurements under PLPL.  (g,h) Position PSDs in the PLPL state for sampling times of 10 and 1,000 s, respectively. (i,j) Corresponding energy PSDs. (k,l) Corresponding force sensitivity spectra. Diamonds mark the loaded weak-force signal peaks and solid lines denote theoretical fits.
}
    \label{fig_PSD}
\end{figure}

To obtain force measurement characteristics of phonon lasers that are closer to practical conditions, a real weak force to be measured is loaded onto the particle. As shown in Fig. \ref{fig_low_power}(a), a 532 nm laser beam illuminates the levitated particle along the X-axis to apply an optical scattering force for measuring. The intensity of this laser is sinusoidally modulated, thereby applying a periodic driving force to the particle. The motion along the X-axis, which serves as the force sensing direction, is set to either a COM cooling state or a PLPL state, respectively, for comparison. The trapping laser power is reduced to $1$ mW during force measurement.

We first test the force measurement characteristics under the COM cooling state.
For a sinusoidal force, $F_{\text{weak}}(t) = F_0 \sin(\Omega_{\mathrm{F}} t)$, its amplitude can be obtained from the position PSD in the cooling state with
    $F_0=\sqrt{S_{xx}(\Omega_{\mathrm{F}})\cdot 4 m^2 \Delta f\left[\left(\Omega_0^2-\Omega_{\mathrm{F}}^2\right)^2+\Omega_{\mathrm{F}}^2\left(\Gamma_0+\delta \Gamma\right)^2\right]}$,
where $S_{xx}(\Omega_{\mathrm{F}})$ is the weak force peak height in the position PSD, $\delta\Gamma$ is the additional cooling damping rate, $\Delta f=1/\tau$ denotes the PSD frequency resolution, and $\tau$ is the sampling time. Thus, for a constant $F_0$, a longer $\tau$ increases $S_{xx}(\Omega_{\mathrm{F}}).$

As shown in Fig. \ref{fig:1}(c), the key drawback for the cooling state force measurement is the long-term drift of the particle's eigen-frequency, $\Omega_0$. 
This instability can be seen in the time-resolved position PSD in Fig. \ref{fig_PSD}(a), where the $\Omega_0$ peak undergoes frequency wandering over the measurement duration. As a consequence, although the short-time position PSD retains a clear Lorentzian profile, as shown in Fig. \ref{fig_PSD}(b), the long-time averaged PSD becomes broadened and split, as shown in Fig. \ref{fig_PSD}(c). This spectral distortion prevents the reliable extraction of force-measurement parameters such as $\Omega_0$ and $\Gamma_0$ from a standard Lorentzian fit over long sampling times. 

Thus, the instability also degrades the frequency-dependent force sensitivity. As shown in Fig. \ref{fig_PSD}(d,e), the force sensitivity spectrum derived from noise analysis exhibits an anomaly as the $\Omega_\text{F}$ approaches the $\Omega_0$. Because the system's response gain is inversely proportional to the frequency difference $\Delta\Omega = |\Omega_0 - \Omega_\text{F}|$, it is extremely sensitive to small drifts in $\Omega_0$, which induce large gain fluctuations. Over time, the accumulated frequency drift smears this gain instability across a wide frequency range, as illustrated in Fig. \ref{fig_PSD}(e).

\begin{figure*}[t]
    \centering
    \includegraphics[width=\linewidth]{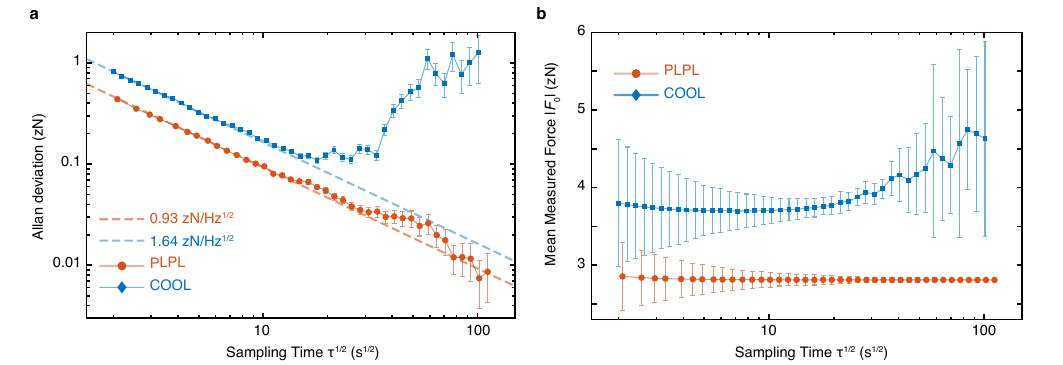}
    \caption{Force sensing performance. (a) Allan deviation of the measured force versus sampling time ($\tau$) for different measurement states. Dashed lines are fits to the $\tau^{1/2}$ dependence, yielding the force sensitivity.  The error bars are estimated using the relative error $\sigma = 1/\sqrt{2(M-1)}$, where $M$ is the effective number of data groups used in the Allan-deviation calculation \cite{Sheimy2008Allan}. (b) Mean value (points) and Allan deviation (error bars) of the measured force versus sampling time. The different measurement states are PLPL (orange circles) and COM cooling (blue squares).
    }
    \label{fig_Allan}
\end{figure*}

The Allan deviation analysis of the measured weak force under COM cooling is shown in Fig. \ref{fig_Allan}(a). 
The noise analysis in Fig. \ref{fig_PSD}(d) predicts a sensitivity of $0.57(4) \mathrm{\times10^{-21}~N/Hz^{1/2}}$. However, instability in the $\Omega_0$ degraded the measured sensitivity to 1.64(11)$\mathrm{\times10^{-21}~N/Hz^{1/2}}$. This instability also limited the optimal averaging time to 360 s, yielding a force resolution about $10^{-22}\mathrm{~N}$. Moreover, Fig. \ref{fig_Allan}(b) shows that exceeding the optimal averaging time leads to increasing error in the estimated force, highlighting the impact of long-term instabilities.

When the phonon laser is used to measure a sinusoidal force, the periodically changing relative phase between them causes the force to periodically perform work on the phonon laser. This results in a periodic change in the energy of the phonon laser, with a frequency $\Delta \Omega$. Therefore, in the energy PSD (e-PSD), a peak can be observed at $\Delta \Omega$, as shown in Fig. \ref{fig:1}(b). The amplitude of the force can be determined from the height of this peak, which is
\begin{equation}
      F_0 = \sqrt{S_{EE}(\Delta \Omega)\cdot 8 m  \Delta f (\Delta \Omega^2 + \gamma_a^2) / |E|},   
\end{equation}
where $S_{EE}(\Delta \Omega)$ is the weak force peak height on the e-PSD.
Therefore, the key to force measurement using phonon lasers is to maintain a stable frequency difference between the force and the phonon laser, making the deployment of PLPL essential. 
The optimal force sensitivity with the phonon laser state is
$S_{\mathrm{F\_PL}} = 2A_{\text{sto}}$ (see Supplementary Information).
Compared to the theoretical sensitivity in the COM cooling state, the force sensitivity in the phonon-laser state is worse by a factor of \( \sqrt{2} \). This is because the measured force signal receives noise contributions from both \( \Omega_0 \pm \Delta \Omega \).

In contrast to the cooling state, the time-resolved position PSD in Fig. \ref{fig_PSD}(f) shows that, under PLPL operation, the carrier peak at the $\Omega_0$ remains frequency and amplitude stable over the entire measurement duration. At the same time, the weak-force signal peak and the corresponding idler peak stay symmetrically distributed about the PLPL main peak. In the position PSDs shown in Fig. \ref{fig_PSD}(g, h), these two sideband peaks can be seen at $\Omega_0 \pm \Delta\Omega$ on the two sides of the carrier peak. In the e-PSD, this signal is combined as a single peak at $\Delta\Omega$, as shown in Fig. \ref{fig_PSD}(i, j). Notably, bumps and several noise peaks present in the position PSD are absent in the e-PSD. This is another advantage of the e-PSD readout, which contains only amplitude noise and is free from phase noise contributions \cite{liu2024nanoscale}. Since the PLPL actively stabilizes the levitation system, the system's force response remains stable during long-term measurements, with the noise floor shape being the same as in short-term results and in agreement with the theoretical fitting. This stability is also shown in the frequency-dependent force sensitivity in Fig. \ref{fig_PSD}(k, l). Even when very close to $\Omega_0$, the force sensitivity remains stable and matches the predictions of the theoretical model. To characterize the system’s performance during actual force measurements, we perform an Allan deviation analysis on the measured force data in Fig. \ref{fig_Allan}(a). The results show that under PLPL, the optimal averaging time for force measurement can reach up to $12,500$ seconds, achieving a force resolution of $8(4)\times10^{-24}$ N at a sensitivity of $9.3(7)\times10^{-22}\mathrm{~N/Hz^{1/2}}$. Furthermore, the measured force values are stable across various measurement durations, as shown in Fig. \ref{fig_Allan}(b).

\section{discussion}

In conclusion, we have demonstrated two distinct applications of phonon laser modes for ultrasensitive force measurements. First, the robust motion signals from the phonon laser enable the automatic updating of feedback parameters, which maintains stable levitation even as trapping power is much reduced. This approach achieves both low-intensity optical trapping and a significant reduction in force noise. Second, the PLPL-based carrier-modulation sensing protocol actively compensates for system instabilities during weak force measurements, extending the optimal averaging time and thus improving the ultimate force resolution.

\begin{figure}[t]
    \centering
    \includegraphics[width=\linewidth]{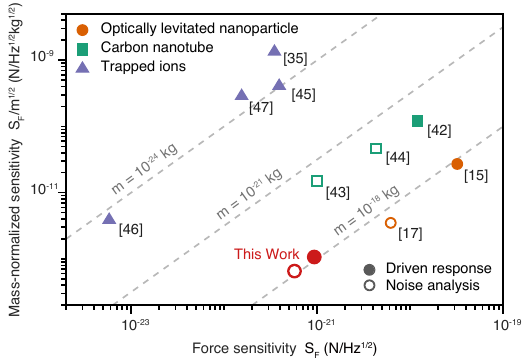}
    \caption{Force sensitivity $S_\mathrm{F}$ versus mass-normalized sensitivity $S_\mathrm{F}/\sqrt{m}$ for optically levitated nanoparticle \cite{hempston2017force,liang2023yoctonewton}, carbon nanotube \cite{Moser2013CNT,Moser2014CNT,Bonis2018CNT} and trapped ions \cite{biercuk2010ultrasensitive,Kevin2021ions,Liu2021Phonon,Bonus2025ions}. Filled symbols denote driven response, in which the sensitivity is obtained from the measured response to an applied force. Open symbols denote noise analysis, in which the sensitivity is inferred from thermal-noise or force-noise analysis. Grey dashed lines are sensor mass guides. Red symbols indicate the present work.
    }
    \label{fig_benchmark}
\end{figure}

A comparison of force sensitivity across different mechanical oscillator sensors is shown in Fig. \ref{fig_benchmark}. For thermally limited mechanical sensors, the force sensitivity is fundamentally bounded by $S_F^{\mathrm{th}}=\sqrt{4 k_{\mathrm{B}} T_0 \Gamma_0 m}$ \cite{hempston2017force,Moser2013CNT,Moser2014CNT}, which generally scales as $\sqrt{m}$. To remove this mass dependence and more directly compare the force-transduction efficiency of different platforms, we consider the mass-normalized sensitivity $S_\mathrm{F}/\sqrt{m}$ in Fig. \ref{fig_benchmark}, together with results from representative optically levitated nanoparticle \cite{hempston2017force,liang2023yoctonewton}, carbon nanotube \cite{Moser2013CNT,Moser2014CNT,Bonis2018CNT}, and trapped-ions \cite{biercuk2010ultrasensitive,Kevin2021ions,Liu2021Phonon,Bonus2025ions}. Under this metric, our system lies the lower bound of the platforms compared here, while reaching an absolute force sensitivity close to values reported for established trapped-ions force sensors.

These results enhance levitated mesoscopic particles as a powerful platform for precision force measurements. Their COM motion directly transduces external perturbations acting on the whole object, while the particle mass, size, material, geometry and charge provide exceptional freedom for engineering the force coupling. This is valuable for applications ranging from electric-field sensing with tuneable charge \cite{moore2014search,frimmer2017controlling,Zhu2023electric} to residual-gas metrology \cite{blakemore2020absolute,liu2024nanoscale,Tseng2025dark} and precision studies of surface and short-range interactions \cite{geraci2010short,blakemore2021nonnewton,Casimir2017,xu2022observation,casimir2023}. It is also attractive for ultralight-dark-matter searches \cite{yin2022dark,kilian2024dark,monteiro2020darkmatter,Tseng2025dark}, where a sensor of suitable size can access whole-particle coherent scattering, strongly enhancing the effective scattering \cite{Afek2022dark}.

Unlike traditional phase control, such as injection locking \cite{Knunz2010locking,pl2024xiao,Kuhn2017rotor} or synchronization \cite{Sheng2020Synch}, which forces the system to oscillate with an internal or external drive while leaving it vulnerable to eigenfrequency drift, the demonstrated PLL phase locks the eigenmode itself, ensuring ultra-long-term sensing stability.
The utility of phase locking extends beyond optical levitation. The technique can be applied to any platform capable of phonon-laser operation, including ion traps and optomechanical systems \cite{pan2024ultra}. Phase locking can improve long-term stability and enhance the performance of existing measurement protocols. Beyond sensing, PLPL may also serve as an actuator for modulating microscopic interactions.
Although the phonon laser is classical rather than quantum in nature, the present low-power trapping approach is also promising for future experiments that require low laser heating and suppression of decoherence. By enabling stable levitation at milliwatt-level optical power, it reduces laser-related decoherence that would otherwise limit coherence in macroscopic quantum control protocols \cite{mq2024bonvin,mq2024lukas,mq2024roda}. It also reduces laser heating of the particle's internal temperature, which is one of the keys for heat-sensitive hybrid systems such as optically levitated nanodiamonds with nitrogen-vacancy centers, where laser heating has remained a key obstacle in high vacuum optical trapping \cite{nd2018conangla2018,nd2018frangeskou2018,nd2020delord,nd2022riviere,nd2024jin}.

\section{Acknowledgements}
This work was supported by the National Natural
Science Foundation of China (Nos. 62225506, 12104438), Chinese Academy
of Sciences Project for Young Scientists in Basic Research (No. YSBR-049), Innovation Program
for Quantum Science and Technology (No. 2021ZD0303200), USTC Major Frontier Research Program (LS2030000002) and the Fundamental Research Funds for the Central Universities. The sample preparation was partially conducted at the
University of Science and Technology of China Center for Micro and Nanoscale
Research and Fabrication.

\bibliography{PL_force}
\end{document}


\title{Supplemental Information: Phase-locked phonon laser enhanced ultra-weak force measurement}

\author{Yu Zheng}
\affiliation{Laboratory of Quantum Information, CAS Center for Excellence in Quantum Information and Quantum Physics, University of Science and Technology of China, Hefei 230026, China}
\author{Long Wang}
\affiliation{Laboratory of Quantum Information, CAS Center for Excellence in Quantum Information and Quantum Physics, University of Science and Technology of China, Hefei 230026, China}
\author{Lyu-Hang Liu}
\affiliation{Laboratory of Quantum Information, CAS Center for Excellence in Quantum Information and Quantum Physics, University of Science and Technology of China, Hefei 230026, China}
\author{Yuan Tian}
\affiliation{Laboratory of Quantum Information, CAS Center for Excellence in Quantum Information and Quantum Physics, University of Science and Technology of China, Hefei 230026, China}
\author{Xiang-Dong Chen}
\affiliation{Laboratory of Quantum Information, CAS Center for Excellence in Quantum Information and Quantum Physics, University of Science and Technology of China, Hefei 230026, China}
\affiliation{Hefei National Laboratory, University of Science and Technology of China,
Hefei 230088, China}

\author{Dong Wu}
\affiliation{State Key Laboratory of Opto-Electronic Information Acquisition and Protection, Key Laboratory of Precision Scientific Instrumentation of Anhui Higher Education Institutes, Department of Precision Machinery and Precision instrumentation, University of Science and technology of China, Hefei 230026, China}

\author{Guang-Can Guo}
\author{Fang-Wen Sun}
 \email{fwsun@ustc.edu.cn}
\affiliation{Laboratory of Quantum Information, CAS Center for Excellence in Quantum Information and Quantum Physics, University of Science and Technology of China, Hefei 230026, China}
\affiliation{Hefei National Laboratory, University of Science and Technology of China,
Hefei 230088, China}

\date{\today}

\maketitle
	\maketitle
	
	\tableofcontents

	\makeatletter
	\renewcommand{\thefigure}{S\@arabic\c@figure}
	\makeatother
	\makeatletter
	\renewcommand{\thetable}{S\@arabic\c@table}
	\makeatother
	\makeatletter
	\renewcommand\thesection{\arabic{section}}
	\renewcommand\thesubsection{\thesection.\arabic{subsection}}
	\makeatother
	
\makeatletter
\renewcommand{\theequation}{S\@arabic\c@equation}
\makeatother
	
	\clearpage

\section{Sensitivity of Force measurement}
This section primarily focuses on the theoretical derivation of the force measurement sensitivity for an optically levitated oscillator, encompassing both the cooling state and the phonon laser state. 
\subsection{Cooling state}
In the cooling state, the equation of motion for a particle subjected to a external force is given by
\begin{equation}
\frac{\mathrm{d}^2 x}{\mathrm{~d} t^2}+\left(\Gamma_0+\delta \Gamma\right) \frac{\mathrm{d} x}{\mathrm{~d} t}+\Omega_0^2 x=\frac{F_{\text {sto }}(t)+F_{\text {weak }}(t)}{m},
\label{cool_dynamic_eq}
\end{equation}
where $x$ is the particle's position, $\Gamma_0$ is the air damping rate, $\delta \Gamma$ is the additional cooling damping rate, $\Omega_0$ is the particle's eigenfrequency, $F_{\text {sto }}(t)=A_{\text {sto }} \zeta(t)$ is the stochastic noise force, and $F_{\text {weak }}(t)=F_0 \cos \left(\Omega_{\mathrm{F}} t\right)$ is the sinusoidal driving force to be measured.

In the following power spectral density (PSD) $S(f)$ analysis, we adopt a two-sided definition, where the total power 
$P$ is given by $P = \int_{-\infty}^{\infty} S(f) \,df = 2 \int_0^{\infty} S(f) \,df$.

Since Eq. (\ref{cool_dynamic_eq}) is a linear differential equation, it can be decomposed into the following two parts, which are solved separately
\begin{equation}
\frac{\mathrm{d}^2 x}{\mathrm{~d} t^2}+\left(\Gamma_0+\delta \Gamma\right) \frac{\mathrm{d} x}{\mathrm{~d} t}+\Omega_0^2 x=\frac{F_{\text {sto }}(t)}{m},
\label{cool_eq_sto}
\end{equation}
and
\begin{equation}
\frac{\mathrm{d}^2 x}{\mathrm{~d} t^2}+\left(\Gamma_0+\delta \Gamma\right) \frac{\mathrm{d} x}{\mathrm{~d} t}+\Omega_0^2 x=\frac{F_{\text {weak }}(t)}{m}.
\label{cool_eq_drive}
\end{equation}

We deal with the driven force first. For a harmonic oscillator driven by a sinusoidal force $F_\text{weak}(t) = F_0 \cos(\Omega_\text{F} t)$, the steady-state displacement response is given by
\begin{equation}
   x_{\text{steady}}(t) = X(\Omega_\text{F}) \cos(\Omega_\text{F} t - \phi) ,
\end{equation}
where the response amplitude $X(\Omega_\text{F})$ and phase shift $\phi$ are, respectively
\begin{equation}
X(\Omega_\text{F}) = \frac{F_0/m}{\sqrt{(\Omega_0^2 - \Omega_\text{F}^2)^2 + ((\Gamma_0+\delta \Gamma) \Omega_\text{F})^2}}, \quad \phi = \tan^{-1}\left(\frac{(\Gamma_0+\delta \Gamma) \Omega_\text{F}}{\Omega_0^2 - \Omega_\text{F}^2}\right).
\end{equation}
The total power, found by integrating the PSD $S(f)$, is equal to the half mean-square displacement
\begin{equation}
\int_0^{\infty} S(f) df = \frac{1}{2}\langle x_{\text{steady}}^2(t) \rangle = \frac{X^2(\Omega_\text{F})}{4}.
\end{equation}
We can have the response function $|\chi(\Omega)|^2$ as
\begin{equation}
|\chi(\Omega)|^2 = \frac{1}{m^2} \frac{1}{(\Omega_0^2 - \Omega^2)^2 + (\Omega (\Gamma_0+\delta \Gamma))^2}.
\end{equation}
Such that the theoretical PSD can be expressed using a delta function:
\begin{equation}
S_\text{weak}(\Omega) = \frac{F_0^2}{4} |\chi(\Omega)|^2 \delta(\Omega - \Omega_\text{F}).
\end{equation}
In a discrete Fourier transform (DFT), this manifests as a sharp peak at the drive frequency $f_\text{F} = \Omega_\text{F} / (2\pi)$. The height of this spectral peak, $S_\text{weak}(f_\text{F})$, is related to the frequency resolution $\Delta f$ of the transform
\begin{equation}
S_\text{weak}(f_\text{F}) = \frac{X^2(\Omega_\text{F})}{4\Delta f}=\frac{F_0^2}{4\Delta f} |\chi(\Omega_\text{F})|^2.
\end{equation}
Therefore, the driving force amplitude $F_0$ can be determined from the measured displacement PSD using the formula
\begin{equation}
F_0 = \sqrt{\frac{4 S_\text{weak}(\Omega_\text{F}) \Delta f }{|\chi(\Omega_\text{F})|^2}}=\sqrt{4 S_\text{weak}(\Omega_\text{F}) \Delta f m^2 \left[ (\Omega_0^2 - \Omega_\text{F}^2)^2 + ((\Gamma_0+\delta \Gamma) \Omega_\text{F})^2 \right]}.
\label{S10}
\end{equation}
 For the stochastic force, $F_\text{noise}=A_\text{sto}\zeta(t)$. Its force PSD is a constant $S_\text{noise}(\Omega) = A_{\text{sto}}^2$, and the force noise induced displacement PSD is 
 \begin{equation}
     S_\text{sto}(\Omega)= S_\text{noise}(\Omega) \cdot |\chi(\Omega)|^2=\frac{A_{\text {sto }}^2}{m^2} \frac{1}{\left(\Omega_0^2-\Omega^2\right)^2+\Omega^2(\Gamma_0+\delta \Gamma)^2}.
     \label{S11}
 \end{equation}

When calculating $F_0$ from the PSD, the measured power at the target force's frequency, $S(\Omega_\text{F})$, is a combined result of the stochastic motion ($S_\text{sto}(\Omega_\text{F})$) and the driven motion ($S_\text{weak}(\Omega_\text{F})$). Since the PSD is derived from the magnitude-squared of the Fourier transform, a proper analysis of $S(\Omega_\text{F})$ must start from the raw Fourier transform of the displacement signal. We therefore define the PSD components in terms of their underlying Fourier transforms: $S(\Omega_\text{F}) = |\mathcal{F}|^2$, $S_\text{weak}(\Omega_\text{F}) = |\mathcal{F}_\text{weak}|^2$, and $S_\text{sto}(\Omega_\text{F}) = |\mathcal{F}_\text{sto}|^2$.

$\mathcal{F}_\text{weak}$ is a constant complex, which is 
\begin{equation}
\mathcal{F}_\text{weak}=\frac{F_0|\chi(\Omega_\text{F})|}{\sqrt{4\Delta f}} e^{i\phi}.
\end{equation}

 $\mathcal{F}_{\text{sto}}$ is a complex random variable whose real and imaginary parts are independent random variables, each following a normal distribution $\mathcal{N}(0, \frac{1}{2}A_{\text{sto}}^2|\chi(\Omega_\text{F})|^2)$.

For the measured spectral power signal,
\begin{equation}
    \begin{aligned}
S(\Omega_\text{F})=|\mathcal{F}|^2&=|\mathcal{F}_\text{weak} + \mathcal{F}_{\text{sto}}|^2 \\&= |\mathcal{F}_\text{weak}|^2 + |\mathcal{F}_{\text{sto}}|^2 + 2(\text{Re}(\mathcal{F}_\text{weak})\text{Re}(\mathcal{F}_{\text{sto}}) + \text{Im}(\mathcal{F}_\text{weak})\text{Im}(\mathcal{F}_{\text{sto}}))\\
    &=|\mathcal{F}_\text{weak}|^2 + |\mathcal{F}_{\text{sto}}|^2 + 2|\mathcal{F}_\text{weak}|Z ,
    \end{aligned}
    \label{S13}
\end{equation}
where $Z = \text{Re}(\mathcal{F}_{\text{sto}})\cos\phi + \text{Im}(\mathcal{F}_{\text{sto}})\sin\phi$. $Z$ following a normal distribution $\mathcal{N}(0, \frac{1}{2}A_{\text{sto}}^2|\chi(\Omega_\text{F})|^2)$.

We now analyze the force sensitivity in two distinct regimes, beginning with the high signal-to-noise ratio (SNR) limit. This regime is defined by the condition $|\mathcal{F}_\text{weak}|^2 \gg \langle|\mathcal{F}_{\text{sto}}|^2\rangle$.
Under this assumption, $|\mathcal{F}_{\text{sto}}|^2$ can be ignored. Eq. (\ref{S13}) simplifies to
\begin{equation}
\begin{aligned}
        S(\Omega_\text{F}) &\simeq |\mathcal{F}_\text{weak}|^2 + 2|\mathcal{F}_0|Z \\&= \frac{F_0^2|\chi(\Omega_\text{F})|^2}{{4\Delta f}} + 2\frac{F_0|\chi(\Omega_\text{F})|}{\sqrt{4\Delta f}}Z.
        \label{S14}
\end{aligned}
\end{equation}
Substituting Eq. (\ref{S14}) into Eq. (\ref{S10}) to calculate the measured force $F_\text{m}$, we have
\begin{equation}
    \begin{aligned}
        F_\text{m} &= \sqrt{F_0^2+2\frac{\sqrt{4\Delta f}F_0}{|\chi(\Omega_\text{F})|}Z}
        \\&\simeq F_0+\frac{\sqrt{4\Delta f}}{|\chi(\Omega_\text{F})|}Z.
    \end{aligned}
\end{equation}
$F_\text{m}$ follows a normal distribution $\mathcal{N}(F_0,2\Delta f A_{\text{sto}}^2)$. The force sensitivity, $S_\text{F}$
 , can be defined as the standard deviation of this measurement to a 1 Hz bandwidth. By setting $\Delta f=1\mathrm{~Hz}$, the sensitivity is therefore
\begin{equation}
    S_\text{F}=\sqrt{2}A_{\text{sto}}.
    \label{S16}
\end{equation}

Another limit regime is when there is no driven force ($F_0=0\mathrm{~N}$) and calculate the measured force with pure noise. We can have
\begin{equation}
    |\mathcal{F}|^2 = |\mathcal{F}_{\text{sto}}|^2 = \text{Re}(\mathcal{F}_{\text{sto}})^2 + \text{Im}(\mathcal{F}_{\text{sto}})^2.
\end{equation}

The `force' calculated from noise spectrum with Eq. (\ref{S10}) is 
\begin{equation}
    F_\text{m}^2 = {\frac{4  \Delta f }{|\chi(\Omega_\text{F})|^2}}|\mathcal{F_\text{sto}}|^2.
\end{equation}
Since $|\mathcal{F}_{\text{sto}}|^2$ is the sum of squares of two independent, zero-mean Gaussian variables, it follows a chi-square distribution which is $|\mathcal{F}_{\text{sto}}|^2\sim \chi^2(2,\frac{1}{2}A_{\text{sto}}^2|\chi(\Omega_\text{F})|^2))$. Thus, $F_\text{m}^2$ also follows chi-square distribution, which would have an expectation value $\sqrt{\mathbb{E}[F_{\text{m}}^2]} = {2 \sqrt{\Delta f}} A_{\text{sto}}$ and a standard deviation ${2 \sqrt{\Delta f}} A_{\text{sto}}$. Both of the values are not equal to the standard deviation of the measured force in the high SNR limit, which is $\sqrt{2\Delta f}A_\text{sto}$. This indicates that when SNR is low during force measurement, the measuring result would be larger than the true value as the noise has a positive expectation value. It should be noted that when calculating its force sensitivity or stochastic force through the noise floor, the square value should be calculated first, then averaged or fitted, and finally square-rooted. According to Jensen's Inequality, $\sqrt{\mathbb{E}[F_{\text{m}}^2]}\geq\mathbb{E}[\sqrt{F_{\text{m}}^2}]$.

It means that the force sensitivity can be extracted directly from the system's noise floor. 
This is accomplished by first defining an equivalent noise strength, $A'_\text{sto}(\Omega)$. With the relation between noise and PSD shown in Eq. (\ref{S11}), we have
\begin{equation}
    A'^2_\text{sto}(\Omega)=\frac{{S(\Omega)}}{|\chi(\Omega)|^2}.
\end{equation}
With Eq. (\ref{S16}), the corresponding frequency-dependent force sensitivity is
\begin{equation}
    S'_\text{F}(\Omega)=\sqrt{2}\sqrt{A'^2_\text{sto}(\Omega)}.
    \label{S20}
\end{equation}

In the scenario where the PSD is entirely dominated by force noise, such that $S(\Omega) \approx S_\text{sto}(\Omega)$, the frequency-dependent sensitivity $S'_\text{F}(\Omega)$ reduces to the constant value $S_\text{F} = \sqrt{2}A_{\text{sto}}$. However, in experimental data, this condition is typically only satisfied near the mechanical eigenfrequency. Away from resonance, measurement noise, primarily shot noise, begin to dominate and become the limiting factor for measurement uncertainty.
 Therefore, Eq. (\ref{S20}) provides a general method for determining the frequency-dependent force sensitivity.

\subsection{Phonon laser state}
To characterize the use of the phonon laser state for ultra-weak force measurement, we must analyze its dynamics under an external driving force. However, in the phonon laser state, the oscillator is subject to a nonlinear modulation, which prevents the separation of the stochastic and external driving force effects within the equation of motion. To address this issue, we instead start from the dynamical equation for energy to calculate the force sensing sensitivity in the phonon laser state.

The phonon laser state is generated by deploy a energy (phonon number) depended damping to the levitated oscillator, which is 
\begin{equation}
    \delta\Gamma(E)=\frac{\gamma_c}{\hbar\Omega_0}E-\gamma_a,
\end{equation}
where $E$ is the mechanical energy of the oscillator, $\gamma_c$ is the non-linear cooling coefficient, $\gamma_a$ is the linear heating coefficient.

The equation for the energy dynamics in the phonon laser state is \cite{pl2023zheng}
\begin{equation}
    d E=\left[-E\left(\Gamma_{\mathrm{0}}+\frac{\gamma_c}{\hbar \Omega_0} E-\gamma_a\right)+\frac{A_\text{sto}^2}{2m}\right] d t+A_\text{sto}\sqrt\frac{E}{m} d W,
\end{equation}

When the oscillator is subjected to a loaded driven force, this force does work on it. The energy equation for the phonon laser, now driven by a periodic external force $F_{\text {weak }}(t)=F_0 \sin \left(\Omega_{\mathrm{F}} t\right)$ becomes
\begin{equation}
    d E=\left[-E\left(\Gamma_{\mathrm{0}}+\frac{\gamma_c}{\hbar \Omega_0} E-\gamma_a\right)+\frac{A_\text{sto}^2}{2m}+\frac{F_0 \sqrt{E}}{\sqrt{2 m}} \sin (\Delta \Omega_\text{F} t+\theta)\right] d t+A_\text{sto}\sqrt\frac{E}{m} d W\text{,}
    \label{PEF1}
\end{equation}
where $\Delta \Omega_\text{F}=\left|\Omega_0-\Omega_{\mathrm{F}}\right|$ is the detuning and $\theta$ is the initial phase difference between the driving force and the particle's oscillation.

In the phonon laser state, the particle's energy exhibits small fluctuations around its mean value, which is $|E|={\gamma_a\hbar\Omega_0}/{\gamma_c}$. We therefore approximate the dynamics by replacing the energy $E$ with its mean value $|E|$, in the Eq. (\ref{PEF1}), yielding
\begin{equation}
    d E=\left[-|E|\left(\Gamma_{\text {0 }}+\frac{\gamma_c}{\hbar \Omega_0} E-\gamma_a\right)+\frac{A_\text{sto}^2}{2m}+\frac{F_0 \sqrt{|E|}}{\sqrt{2 m}} \sin (\Delta \Omega_\text{F} t+\theta)\right] d t+A_\text{sto}\sqrt\frac{|E|}{m} d W\text{.}
    \label{PEF2}
\end{equation}
We now decompose Eq. (\ref{PEF2}) into its stochastic and driven components. The energy equation for the stochastic component is
\begin{equation}
    d E=\left[-|E|\left(\Gamma_{\mathrm{0}}+\frac{\gamma_c}{\hbar \Omega_0} E-\gamma_a\right)+\frac{A_\text{sto}^2}{2m}\right] d t+A_\text{sto}\sqrt\frac{|E|}{m} d W\text{.}
    \label{PEF3}
\end{equation}
Applying a Fourier transform to Eq. (\ref{PEF3}) and neglecting the DC component, we obtain
\begin{equation}
    -i \Omega \tilde{E}=-|E| \frac{\gamma_c}{\hbar \Omega_0} \tilde{E}+A_\text{sto}\sqrt\frac{|E|}{m} \tilde{W}\text{.}
    \label{PEF4}
\end{equation}
The corresponding energy PSD is
\begin{equation}
    S^{EE}_{\text {sto }}(\Omega)=\frac{A_\text{sto}^2}{m} \frac{|E|}{\Omega^2+\gamma_a^2}\text{.}
    \label{PEF5}
\end{equation}

Now, we solve the driven part. The energy equation for the driven component is
\begin{equation}
    d E=\left[-|E|\left(\Gamma_{\mathrm{0}}+\frac{\gamma_c}{\hbar \Omega_0} E-\gamma_a\right)+\frac{F_0 \sqrt{|E|}}{\sqrt{2 m}} \sin (\Delta \Omega_\text{F} t+\theta)\right] d t\text{.}
    \label{PEF6}
\end{equation}

Similarly, applying a Fourier transform to Eq. (\ref{PEF6}) and neglecting the DC component gives
\begin{equation}
    -i \Omega \tilde{E}=-|E| \frac{\gamma_c}{\hbar \Omega_0} \tilde{E}+\frac{F_0 \sqrt{|E|}}{\sqrt{2 m}} \frac{1}{2} \delta(\Omega-|\Delta \Omega_\text{F}|)\text{.}
    \label{PEF7}
\end{equation}

The corresponding PSD for the driven signal is
\begin{equation}
S^{EE}_{\text {weak }}(\Omega)=\frac{F_0^2}{8 m } \frac{|E|}{\Omega^2+\gamma_a^2} \delta(\Omega-|\Delta \Omega_\text{F}|)\text{.}
\end{equation}

In a DFT, this manifests as a sharp peak at the $\Delta \Omega_\text{F}$. The height of this spectral peak, $S^{EE}_\text{weak}(\Delta \Omega_\text{F})$, is related to the frequency resolution $\Delta f$ of the transform
\begin{equation}
S^{EE}_{\text {weak }}(\Delta\Omega_\text{F})=\frac{F_0^2}{8 m \Delta f} \frac{|E|}{\Delta\Omega_\text{F}^2+\gamma_a^2} \text{.}
\end{equation}
The driven force can be calculated with
\begin{equation}
    F_0=\sqrt{8S^{EE}_{\text {weak }}(\Delta\Omega_\text{F})m \Delta f(\Delta \Omega_\text{F}^2+\gamma_a^2)/|E|}\text{.}
\end{equation}
Similar to the force sensitivity discussion in the cooling state. The force sensitivity of phonon laser under high SNR condition is
\begin{equation}
    S_\text{F\_PL}=2A_\text{sto}.
\end{equation}
The frequency depended force sensitivity is
\begin{equation}
    S'_\text{F\_P}(\Omega)=2\sqrt{m|S^{EE}(\Omega)|(\Omega^2+\gamma_a^2)/|E|}.
\end{equation}

\section{Experiment Setup}

\begin{figure*}[t]
    \centering
    \includegraphics[width=\linewidth]{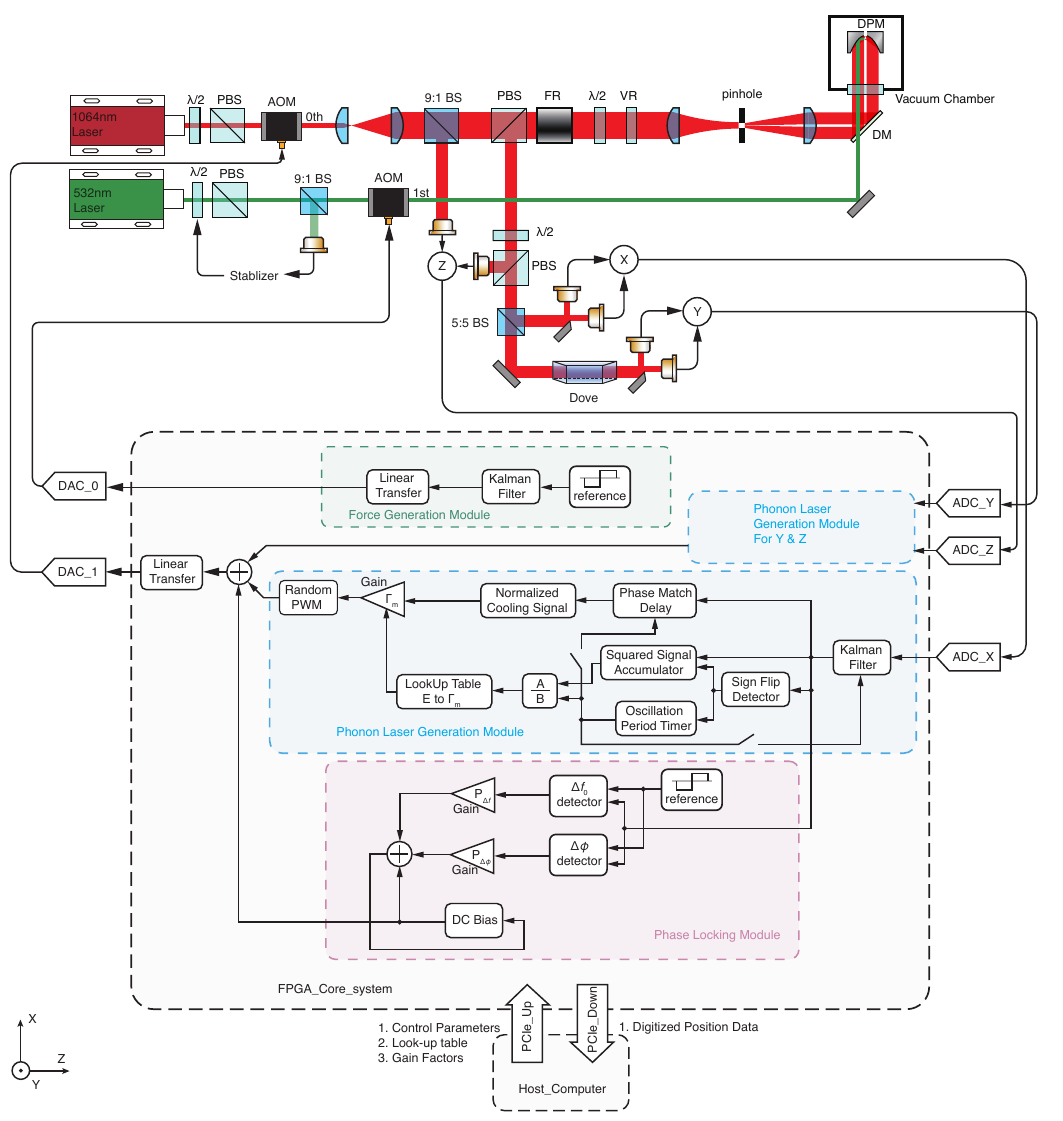}
    \caption{Schematic diagram of an experimental setup including optics and electronics
    \label{Exp_set}}
\end{figure*}

The experimental setup is depicted in Fig. \ref{Exp_set}.

\subsection{Optical setup}

A continuous-wave (CW) 1064~nm laser serves as the trapping beam. 
The laser first passes through an intensity modulator consisting of a half-wave plate (HWP, marked as $\lambda/2$ in Fig. \ref{Exp_set}) mounted on a motorized rotator and a polarizing beam splitter (PBS), enabling large range control of the optical power. 
The beam is then sent through an acousto-optic modulator (AOM), where the zeroth-order diffracted beam is selected for trapping.

After the AOM, the beam is expanded to a diameter of 7~mm using a telescope and subsequently split by a 9:1 beam splitter (BS). 
The transmitted beam is used for trapping, whereas the weak reflected beam provides a power reference for balanced detection of the particle motion along the $z$ axis.

In the trapping path, the beam passes through a PBS and a Faraday rotator (FR), followed by a HWP and a m = 1 vortex retarder (VR). 
The HWP rotates the incident polarization such that the VR generates a radially polarized beam rather than an azimuthally polarized one.

Immediately after the VR, the field corresponds to a radially polarized Gaussian beam. 
A pinhole spatial filter removes higher-order components and produces a 
 radially polarized doughnut (RPD) beam \cite{dorn2003sharper, novotny2012principles}. 
The RPD beam is directed onto a deep parabolic mirror (DPM), which tightly focuses the field and generates a strongly focal field with polarization along the $z$ direction for optical trapping.

Both the trapping light transmitted through the focus and the light scattered by the trapped particle are reflected and collimated by the DPM  for backward position detection \cite{magrini2021real,tebbenjohanns2021quantum}. 
The returning beam passes through the VR again, where the radial polarization is converted back to linear polarization. 
After passing once more through the HWP plate and the FR, the beam is reflected by the PBS and directed to the detection system for three-dimensional position readout.

A separate CW 532 nm laser is used to apply weak external forces. Its intensity is controlled via an HWP and PBS. A 9:1 beam splitter (BS) taps off a portion of the beam for a custom feedback loop that stabilizes the laser intensity, ensuring long-term force stability. The first-order diffracted beam from the AOM is used to apply a sinusoidally modulated force to the levitated particle along the $x$ axis.

The parabolic mirror surface is described in polar coordinates by
\begin{equation}
R_{\mathrm{inc}}=\frac{2f\sin\theta}{1+\cos\theta}.
\end{equation}

In our experiment, the entrance pupil of the mirror is restricted to 
$R_{\mathrm{inc}}\in[0.5f,\,5f]$, where the focal length is $f=1\,\mathrm{mm}$. 
This corresponds to a focusing angle
\begin{equation}
\theta = 2\arctan\left(\frac{R_{\mathrm{inc}}}{2f}\right)
\in [28^{\circ},\,136^{\circ}].
\end{equation}

To minimize the focal spot size, the peak-intensity ring radius of the incident RPD beam is chosen as $w_0=2.3f$ 
\cite{salakhutdinov2016optical, sondermann2011maximizing}. There is a 1 mm diameter hole on the top of the DPM for particle delivery.

\subsection{FPGA-based control system}

The weak force signal generation, feedback motion control, and phonon laser phase locking are implemented on a field-programmable gate array (FPGA). 

For the weak force signal. To generate a pure sinusoidal force, a reference square wave is first derived by frequency division from the 50 MHz master clock. A Kalman filter then synthesizes a pure sinusoidal signal from this square wave, to which a DC offset is added to ensure the final signal is always positive. We chose this method over Direct Digital Synthesis (DDS) because DDS still has phase noise introduced spectral broadening that would be observable in long-term measurements. To correct for the nonlinear relationship between AOM control voltage and diffraction efficiency, the synthesized sine wave is passed through a linearization lookup table (LUT) before being sent to the AOM controller via a digital-to-analog converter (DAC).

For the phonon laser and COM cooling, the control logic is identical for all three axes. The motion control is based on the parametric feedback control \cite{paraCool2012}. Position signals from the analog-to-digital converter (ADC) are first filtered by a Kalman filter. The signal then enters a phase-match delay module that compensates for loop delays and generates in-phase (I) and quadrature (Q) signals by adding additional delay. The product of these two signals, followed by an extraction of the sign bit, yields a normalized cooling signal.

The phonon laser requires the feedback gain to be dependent on the oscillator's energy. To achieve this, the squared position signal is integrated over each oscillation cycle. A oscillation period timer that trigged by a sign flip detection determines the oscillation period and frequency. The particle's motional energy, proportional to this integrated value divided by the period, is thus calculated in real-time. This energy value then addresses an LUT that maps motional energy to the feedback damping gain. The contents of this LUT are pre-calculated on a host computer and uploaded to the FPGA. The COM cooling can be achieved by uploading a LUT with identical values. The final feedback signal is the product of this feedback damping gain and the normalized cooling signal. 

Under precision measurement conditions, the feedback itself can be a noise source, requiring minimal modulation damping. We found that our AOM could not reliably generate modulation depths below 0.003\% (equivalent to a ~10 Hz damping rate for a 150 kHz oscillator), while our experiments required damping rates below 1 Hz. To overcome this limitation, we implemented a random pulse-width modulation (PWM) scheme. For feedback damping gain below the threshold, the random PWM will be engaged. The amplitude of output pulses are held constant, and the effective damping is controlled by varying the duty cycle of the output pulses. The pulses are applied at random intervals to avoid introducing spurious spectral peaks. This method, combined with 32-bit control precision, theoretically allows for modulation depths as low as $10^{-9}$\% ($\sim3 \mathrm{~\mu Hz}$ damping rate). The 3-axes motion control signals are combined and added with the phase locking signal.

The Kalman filters and phase-matching modules require accurate frequency and period parameters. While these can be updated in real-time from the oscillation period timer, this is only safe in the phonon laser state. In standard cooling mode, the low SNR position signal makes real-time updates unreliable. Noise can lead to occasionally incorrect period calculations, causing the feedback damping to become negative (heating) and resulting in particle loss. Given that recovering from a particle loss event have to re-pump the system from atmosphere to high vacuum, which takes over a week. This risk is unacceptable. Conversely, in the phonon laser state, the high SNR of the oscillation enables accurate and stable real-time frequency tracking, which is essential for maintaining control as the trapping power is adjusted.

In the phase-locking module, we modulate the phonon laser's frequency and phase by varying the trapping laser power via the AOM's DC bias. In this scheme, the particle's phonon laser signal is compared to a reference signal, generating both a frequency error and a phase error. These two error signals are scaled by independent gains and summed. This combined correction signal is then fed into a digital integrator, which update the DC-bias per oscillation cycle, triggered by the rising edge of the particle's position signal. The final integrated DC-bias value is added to the parametric feedback control signal, and sent through a linearization lookup table (LUT) before being sent to the AOM controller via a digital-to-analog converter (DAC).

When setting the parameters of the phase-locking module, the following principles are mainly adopted. For the frequency-error gain coefficient, an optimal value can be calculated from the relationship between the frequency-error gain coefficient and the frequency variation of the levitated particle. For the phase-error gain coefficient, a smaller value leads to smaller side bumps around the main peak in the spectrum. However, it must also remain sufficiently large to suppress substantial phase excursions and thereby prevent phase unlock. Another key parameter affecting the phase-locking performance is the Kalman gain of the Kalman filter. There exists an optimal Kalman gain that minimizes the phase error in the locked state.

\section{Experimental Parameters}
This chapter mainly introduces the experimental parameter calibrations and settings used during the experiments.

\subsection{System calibration and error analysis}

To convert the measured voltage signal into an absolute displacement and to determine the particle mass used in force calibration, we carry out a thermomechanical calibration based on the equipartition theorem and the gas-damping model of a spherical particle in the Epstein regime \cite{Ricci2019accurate,zheng2020robust}. Since the subsequent force extraction depends explicitly on both the displacement calibration factor and the particle mass, the uncertainty of this calibration forms an important part of the systematic error budget of the measured force.

After a particle is loaded into the optical trap, the chamber pressure is first reduced below $10^{-6}$ mbar and kept there for a while in order to remove volatile adsorbates on the particle surface or trapped inside the particle. The pressure is then increased to the range of $10$--$50$ mbar and held constant. In this pressure range, the gas damping is sufficiently strong to allow reliable fitting of the thermomechanical PSD. From the fitted PSD, the mean-square voltage, eigen-frequency, and damping rate are extracted. The particle mass and diameter are then inferred from the gas-damping model, and the voltage-to-displacement calibration factor is obtained from the equipartition relation.

For the particle used in the force measurement reported in this work, the thermal calibration is performed at a pressure of $14.4$ mbar and environment temperature $298$ K. For the sensing axis, the fitted parameters are
\begin{equation}
\langle V_x^2\rangle = 0.02999~\mathrm{V^2},\qquad
\Omega_{0x}/2\pi=4.625\times10^5~\mathrm{Hz},\qquad
{\Gamma_{0x}}/{2\pi}=1.884\times10^4~\mathrm{Hz}.
\end{equation}

For a spherical particle in the molecular-flow regime, the particle radius can be obtained from the Epstein damping formula as
\begin{equation}
R=
\frac{p}{\rho \Gamma_0}
\sqrt{\frac{\pi m_g}{2k_B T}}
\left(1+\frac{\pi a}{8}\right),
\label{S_cal_radius}
\end{equation}
where $\rho$ is the particle density, $p$ is the gas pressure, $T$ is the gas temperature, $m_g$ is the molecular mass of the background gas, and $a$ is the momentum accommodation coefficient. Combining Eq. (\ref{S_cal_radius}) with
\begin{equation}
m=\frac{4}{3}\pi R^3\rho,
\label{S_cal_mass_radius}
\end{equation}
the particle mass can be written as
\begin{equation}
m=
\frac{4}{3}\pi \rho
\left[
\frac{p}{\rho \Gamma_0}
\sqrt{\frac{\pi m_g}{2k_B T}}
\left(1+\frac{\pi a}{8}\right)
\right]^3.
\label{S_cal_mass_full}
\end{equation}
Therefore, the damping-based mass calibration follows the scaling relation
\begin{equation}
m \propto \rho^{-2}\Gamma_0^{-3}p^{3}T^{-3/2}\left(1+\frac{\pi a}{8}\right)^3.
\label{S_cal_mscale}
\end{equation}

Using the fitted thermal parameters, the calibrated particle properties are
\begin{equation}
m = 7.599\times10^{-19}~\mathrm{kg},
\qquad
d = 89.86~\mathrm{nm}.
\end{equation}

The voltage-to-displacement conversion coefficient is obtained from the equipartition relation
\begin{equation}
m=\frac{k_B T}{c_x^2\langle V_x^2\rangle \Omega_{0x}^2},
\end{equation}
which gives
\begin{equation}
c_x=\sqrt{\frac{k_B T}{m\langle V_x^2\rangle \Omega_{0x}^2}}.
\label{S_cal_cx}
\end{equation}
Using the calibrated mass and the fitted thermal parameters, we obtain
\begin{equation}
c_x = 146.2~\mathrm{nm/V}.
\end{equation}

All parameters and uncertainties used in the calibration are summarized in Table \ref{tab_calib_error}. 

\begin{table}[htbp]
\centering
\caption{Parameters and relative uncertainties used in the system calibration and force-error analysis.}
\begin{tabular}{lcc}
\hline
Parameter & Value & Relative uncertainty \\
\hline
Pressure $p$ & 14.4 mbar & $0.2\%$ \\
Temperature $T$ & 298 K & $0.4\%$ \\
Mean-square voltage $\langle V_x^2\rangle$ & 0.02999 V$^2$ & $0.5\%$ \\
Mechanical eigen-frequency $\Omega_{0x}/2\pi$ & 462.5 kHz & $1\%$ \\
Gas damping rate $\Gamma_{0x}/2\pi$ & 18.8 kHz & $1\%$ \\
Particle density $\rho$ & $2000~\mathrm{kg/m^3}$ & $5\%$ \\
Accommodation coefficient $a$ & $0.9$ & $11.1\%$ \\
Feedback-induced damping $\delta\Gamma$ & 10 Hz & $20\%$ \\
\hline
Particle mass $m$ & $7.599\times10^{-19}$ kg & $13.6\%$ \\
Calibration factor $c_x$ & 146.2 nm/V & $6.9\%$ \\
\hline
\end{tabular}
\label{tab_calib_error}
\end{table}

Using Eq. (\ref{S_cal_mscale}), the relative uncertainty of the particle mass is
\begin{equation}
\left(\frac{\sigma_m}{m}\right)^2
=
\left(2\frac{\sigma_\rho}{\rho}\right)^2
+
\left(3\frac{\sigma_{\Gamma_0}}{\Gamma_0}\right)^2
+
\left(3\frac{\sigma_p}{p}\right)^2
+
\left(\frac{3}{2}\frac{\sigma_T}{T}\right)^2
+
\left[
3\frac{\pi a/8}{1+\pi a/8}\frac{\sigma_a}{a}
\right]^2,
\label{S_cal_merr}
\end{equation}
which gives
\begin{equation}
\frac{\sigma_m}{m}\approx 13.6\%.
\label{S_cal_mfinal}
\end{equation}

Since $c_x\propto m^{-1/2}$, the relative uncertainty of the displacement calibration factor is
\begin{equation}
\left(\frac{\sigma_{c_x}}{c_x}\right)^2
=
\left(\frac12\frac{\sigma_T}{T}\right)^2
+
\left(\frac12\frac{\sigma_m}{m}\right)^2
+
\left(\frac12\frac{\sigma_{\langle V_x^2\rangle}}{\langle V_x^2\rangle}\right)^2
+
\left(\frac{\sigma_{\Omega_{0x}}}{\Omega_{0x}}\right)^2,
\label{S_cal_cxerr}
\end{equation}
which gives
\begin{equation}
\frac{\sigma_{c_x}}{c_x}\approx 6.9\%.
\label{S_cal_cxfinal}
\end{equation}

In the COM-cooling state, the force amplitude is extracted from the displacement PSD peak according to Eq. (\ref{S10}),
\begin{equation}
F_0=
\sqrt{
4 S_{xx}(\Omega_\mathrm{F}) \Delta f\, m^2
\left[
(\Omega_0^2-\Omega_\mathrm{F}^2)^2+\Omega_\mathrm{F}^2(\Gamma_0+\delta\Gamma)^2
\right]
}.
\label{S_force_raw}
\end{equation}
Here $S_{xx}$ is not an independently calibrated quantity. Since the displacement is obtained from the voltage readout through
\begin{equation}
x=c_xV_x,
\end{equation}
the corresponding PSD satisfies
\begin{equation}
S_{xx}=c_x^2S_{VV}.
\end{equation}
At the same time, the calibration factor $c_x$ is itself determined from the calibrated mass through Eq. (\ref{S_cal_cx}). Therefore, the uncertainties of $S_{xx}$ and $m$ are correlated and must not be counted independently.

Substituting
\begin{equation}
c_x^2=\frac{k_B T}{m\langle V_x^2\rangle \Omega_0^2}
\end{equation}
into Eq. (\ref{S_force_raw}), we obtain the equivalent force-calibration formula
\begin{equation}
F_0=
\sqrt{
4\Delta f\,S_{VV}
\frac{k_B T\, m}{\langle V_x^2\rangle\Omega_0^2}
\left[
(\Omega_0^2-\Omega_\mathrm{F}^2)^2+\Omega_\mathrm{F}^2(\Gamma_0+\delta\Gamma)^2
\right]
}.
\label{S_force_corr}
\end{equation}

Accordingly, the relative uncertainty of the extracted force is
\begin{equation}
\left(\frac{\sigma_{F_0}}{F_0}\right)^2
\approx
\left(\frac12\frac{\sigma_m}{m}\right)^2
+
\left(\frac12\frac{\sigma_T}{T}\right)^2
+
\left(\frac12\frac{\sigma_{\langle V_x^2\rangle}}{\langle V_x^2\rangle}\right)^2
+
\frac14\left(\frac{\sigma_A}{A}\right)^2,
\label{S_force_err_used}
\end{equation}
where
\begin{equation}
A=(\Omega_0^2-\Omega_\mathrm{F}^2)^2+\Omega_\mathrm{F}^2(\Gamma_0+\delta\Gamma)^2.
\end{equation}

Under the present experimental conditions, the force frequency is detuned from the mechanical eigenfrequency by about $1\%$, while the uncertainties of $\Omega_0$ and $\Omega_\mathrm{F}$ in the final force extraction are negligible. In this case, the response factor $A$ is dominated by the detuning term, and the additional uncertainty associated with $\delta\Gamma$ gives only a small correction. Therefore, the total systematic uncertainty of the force calibration is well approximated by the calibration-related baseline uncertainty,
\begin{equation}
\frac{\sigma_{F_0}}{F_0}
\approx
\sqrt{
\left(\frac12\frac{\sigma_m}{m}\right)^2
+
\left(\frac12\frac{\sigma_T}{T}\right)^2
+
\left(\frac12\frac{\sigma_{\langle V_x^2\rangle}}{\langle V_x^2\rangle}\right)^2
}
\approx 6.8\%.
\label{S_force_err_final}
\end{equation}
Thus, in the present force measurement configuration, the uncertainty of the measured force is mainly inherited from the mass calibration. The same conclusion also applies to the error analysis of the phonon-laser force measurement and force noise, yielding an identical uncertainty of 6.8\% as in the COM cooling force measurement.

\subsection{Laser power reduction}

\begin{figure*}[t]
    \centering
    \includegraphics[width=\linewidth]{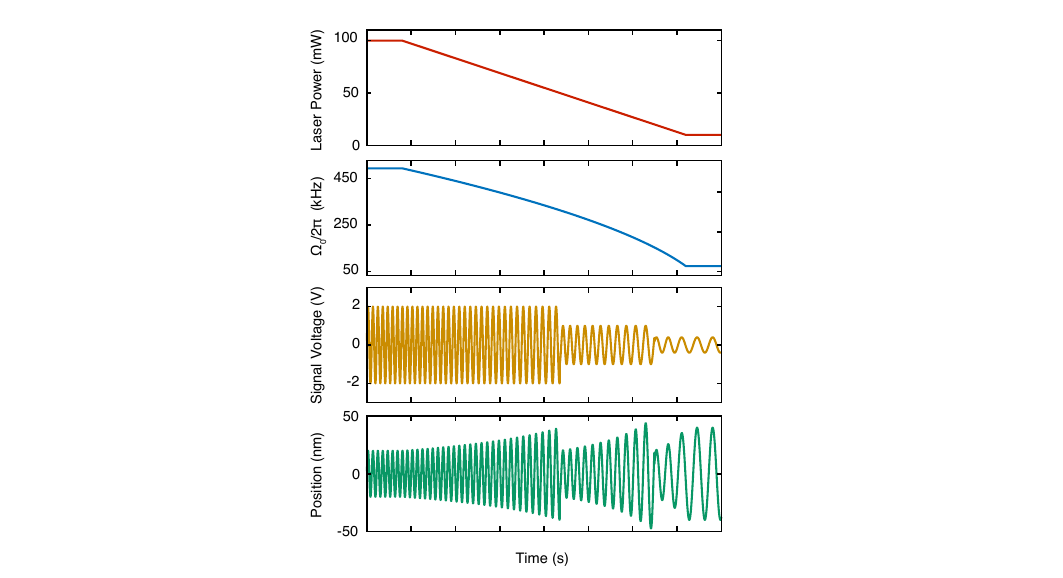}
    \caption{Schematic of the trapping laser power reduction procedure. As the laser power is reduced, the oscillation eigen-frequency is decreasing. The phonon laser modulation keep the oscillator's voltage signal amplitude constant. To maintain stable levitation, the phonon laser parameters are occasionally reset to keep the particle's actual amplitude within the safe limit.
    \label{fig_low_power}}
\end{figure*}

Several parameters scale with the incident trapping power ($P_\text{inc}$), which can be inferred from the measured trap frequency ($\Omega_0$). The key relationships are:
\begin{align}
       \Omega_0 \propto \sqrt{P_\text{inc}}\text{.}\\
   c_\text{calib} \propto 1/P_\text{inc}\text{.}
\end{align}

 To ensure consistent feedback damping across different power levels, the conversion factor between modulation depth ($\eta$) and the resulting damping ($\delta\Gamma$), defined as $c_\text{M} = \delta\Gamma/\eta$, must also be adjusted occasionally, following the relation
 \begin{equation}
     c_\text{M} \propto P_\text{inc} \text{.}
 \end{equation}

As illustrated in Fig. \ref{fig_low_power}, the power reduction protocol proceeds as follows: At a pressure about $2\times10^{-9}$ mbar, free-run phonon laser control is applied along all three axes (X, Y, Z). Subsequently, a frequency-tracking module is activated to automatically update period and frequency parameters for the feedback control. Before power reduction, the BPD gain is increased to $500$ kV/A so that the photon shot noise can exceed the electronic noise, and the voltage signal wouldn't be too small under 1 mW trapping power. During power reduction, we occasionally adjust the control parameters of the phonon laser to decrease the phonon laser's amplitude, so that the real amplitude of the phonon laser remains within a safe range.

 \subsection{Feedback settings for force measurements}
 
During COM cooling, a cooling damping of $\delta\Gamma=10\mathrm{~Hz}$ is applied to the X-axis, phonon laser control with $\gamma_c=10^{-5}\mathrm{~Hz}$ and $\gamma_a=0\mathrm{~Hz}$ is applied to the Y-axis, and phonon laser control with $\gamma_c=2\times 10^{-6}\mathrm{~Hz}$ and $\gamma_a=0\mathrm{~Hz}$ is applied to the Z-axis. Under this control state, the amplitudes of the Y and Z axes are cooled while minimizing the feedback damping on the Y and Z axes as much as possible. This both avoids cross-coupling of YZ-axis motion to X-axis motion and minimizes the impact of feedback control on force measurement performance.

During phonon lasing force measurement, the Y and Z axes are controlled in the same way as during COM cooling, while the X-axis is subjected to phonon lasing control with $\gamma_c=2\times10^{-6}\mathrm{~Hz}$ and $\gamma_a=1\mathrm{~Hz}$. This minimizes the impact of feedback control on the force measurement signal as much as possible.

It should also be noted that the Kalman filter also affects the results. A larger Kalman factor provides a narrower filter bandwidth, which can filter out more noise. This is meaningful when pursuing the ultimate cooling temperature. However, for force measurement, an excessively narrow filter bandwidth will alter the uniformity of the feedback control’s frequency-domain response, making it difficult to fit the frequency-domain signal to obtain force noise or force sensitivity.

\bibliography{PL_force}